\documentclass[preprint,showpacs,preprintnumbers,amsmath,amssymb]{revtex4}
\usepackage{amssymb}
\usepackage{graphicx}
\usepackage{dcolumn}
\usepackage{bm}

\begin{document}

\title{Selective Spin Injection Controlled by Electrical way in Ferromagnet/Quantum Dot/Semiconductor system}
\author{Zhen-Gang Zhu}
\affiliation{Center for Advanced Study, Tsinghua University, Beijing
100084, China;\\
Institut f\"{u}r Physik,
Martin-Luther-Universit\"{a}t Halle-Wittenberg,
Nanotechnikum-Weinberg, Heinrich-Damerow-St. 4,
06120 Halle, Germany;\\
Silicon Nano Device Lab (SNDL), ECE Department, National University
of Singapore; }

\begin{abstract}
Selective and large polarization of current injected into
semiconductor (SC) is predicted in Ferromagnet (FM)/Quantum Dot
(QD)/SC system by varying the gate voltage above the Kondo
temperature. In addition, spin-dependent Kondo effect is also
revealed below Kondo temperature. It is found that Kondo resonances
for up spin state is suppressed with increasing of the polarization
$P$ of the FM lead. While the down one is enhanced. The Kondo peak
for up spin is disappear at $P=1$.
\end{abstract}

 \maketitle

Effective spin injection into semiconductor is the central issue of
spin-related semiconductor devices, such as the so-called
spin-field-effect transistor (SFFT) proposed by Datta and Das
\cite{datta} which may be the original starting point of spintronics
\cite{wolf}. Spin-valve effect was predicted in it via controlling
the gate voltage which controls the Rashba spin-orbit coupling
parameter \cite{datta}. Some experimental attempts were then
performed to realize it but only small signal of spin injection had
been observed. Schmidt \textit{et al.} pointed out that the mismatch
of conductance of FM and SC is the reason of the low efficiency of
spin injection \cite{schmidt1}. However, Rashba proposed that a
tunnel barrier can be inserted between the FM and SC to overcome
this problem \cite{rashba}. Soon, many experiments were then
reported to confirm Rashba's idea \cite{mtt,tb}. For example, hot
electron current with a high spin polarization of about $98\%$ can
be obtained \cite{mtt}. On the other hand, other methods for spin
filter or spin injection into semiconductor are also proposed, such
as a FM tip of scanning tunnelling microscope is used to inject
spin-polarized electrons into SC \cite{stm} and a triple tunnel
barrier diode is utilized as spin source to enhance the
spin-filtering efficiency even to $99.9\%$ \cite{koga}.

More recently, new attempts to realize the devices where the spin
character of the injected and detected electrons could be voltage
selected \cite{gruber}\cite{slobodskyy}, have been made. In these
devices, the source-drain voltage-controlled spin filter effect is
investigated in a magnetic resonant tunnelling diode structure in
which the central spacer is made of dilute magnetic SC ZnMnSe. Zhu
and Su \cite{zhu} proposed a magnetic filed dependence spin filter
effect based on ZnSe/ZnMnSe/ZnSe/ZnMnSe/ZnSe structure in which
resonances of different spin components occur at different magnitude
of magnetic field. These researches open new ways to controllable
spin filter effect. However, these proposed structures are involved
in dilute magnetic SC whose Curie temperature is known blow room
temperature, preventing its further application in devices. In
addition, for the difficulty of operating individual spin by
external magnetic field, new attempt called all electrical devices
is proposed in which the controlling are all via electrical ways.

In this letter, such a selective spin injection into semiconductor
is predicted in Ferromagnet (FM)/Quantum Dot (QD)/SC system by
varying the gate voltage which controls the states of the QD. A FM
layer holding high Curie temperature (above room temperature) is
used as a spin source and polarized electrons flowing out of it
tunnel through a vertical QD (VQD) \cite{vqd} into SC. Between the
two tunnel barrier a quantum well is defined as a QD with strong
Coulomb interaction. The energy levels of QD can be tunned by a gate
voltage $V_{g}$. It is found the polarization of current is large
and can be controlled by tunning $V_{g}$ from negative to positive
(from down-spin filtering to up-spin filtering) because of the mixed
roles of Coulomb interaction and the splitting of spin subbands of
FM. It is worth pointing out that the splitting of energy levels of
QD for different spins are large and corresponds to the Curie
temperature order. This large splitting guarantees the well-defined
separation of polarized current with different spins and the spin
filter effect.

%%%%%%%%%%%%%%%%%%%%%%%%%%%%%%%%%%%%%%%%%%%%%%%%%%%%%%%%%%
\begin{figure}[tbph]
\centering \includegraphics[width =9 cm, height=8 cm]{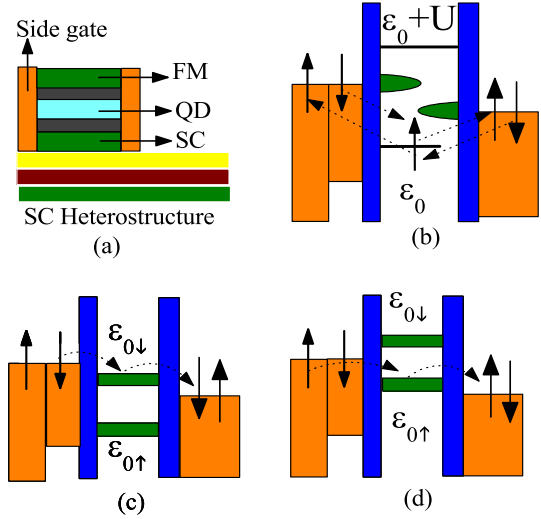}
\caption{(color online). (a) The model configuration. (b) The
formation of spin-dependent Kondo resonances for spin-polarized
lead. (c) and (d) The developed spin-dependent resonant tunnelling.
$\varepsilon_{0\uparrow }$ and $\varepsilon_{0\downarrow }$ have
finite width for the imaginary of selfenergy. } \label{fig1}
\end{figure}
%%%%%%%%%%%%%%%%%%%%%%%%%%%%%%%%%%%%%%%%%%%%%%%%%%%%%%%%
The Hamiltonian is $H=H_{leads}+H_{dot}+H_{T}$,
$H_{leads}=\sum_{k\sigma }\varepsilon_{k\sigma }^{L}a_{k\sigma
}^{\dagger }a_{k\sigma }+\sum_{q\sigma }\varepsilon _{qR}b_{q\sigma
}^{\dagger }b_{q\sigma }$, $H_{dot}=\sum_{\sigma }\epsilon
_{0}d_{\sigma }^{\dagger }d_{\sigma }+Un_{d\uparrow }n_{d\downarrow
}$, $H_{T}=\sum_{k\sigma }[t_{kL}^{\sigma }a_{k\sigma }^{\dagger
}d_{\sigma }+h.c.]+\sum_{q\sigma }[t_{qR}^{\sigma }b_{q\sigma
}^{\dagger }d_{\sigma }+h.c.]$, where $\varepsilon_{k\sigma
}^{L}=\varepsilon_{kL}-\mu_{L}-\sigma M$, $M=g\mu_{B}h/2$, $g$ is
Land\'{e} factor, $\mu _{B}$ is Bohr magneton, $h$ is the molecular
field, $\varepsilon_{kL}$ is the single-particle dispersion of the
left FM, $\mu_{L(R)}$ is the Fermi level of the left (right) lead,
$n_{d\sigma }=d_{\sigma }^{\dagger }d_{\sigma }$,
$\varepsilon_{qR}=\hbar^{2}q^{2}/2m^{\ast }$, $m^{\ast }$ is
effective mass of electrons in the right lead, $t_{kL(qR)}^{\sigma}$
denotes the tunnelling amplitude through the left (right) barrier.

Then following the standard equation of motion method, and assuming
that higher-order spin-correlations in the leads can be neglected
\cite{assum}, the Green function $\langle\langle
d_{\sigma}|d_{\sigma'}^{\dagger }\rangle\rangle^{r}$ can be obtained
\begin{equation}
\langle\langle d_{\sigma}|d_{\sigma^{\prime }}^{\dagger }\rangle
\rangle^{r}=\frac{(\varepsilon-\widetilde{\epsilon }_{\sigma
}+U\left\langle n_{\overline{\sigma }}\right\rangle )\delta_{\sigma
\sigma^{\prime }}-U\left\langle d_{\overline{\sigma }}^{\dagger
}d_{\sigma }\right\rangle\delta_{\overline{\sigma }\sigma^{\prime
}}}{(\varepsilon-\widetilde{\epsilon }_{\sigma })(\varepsilon
-\epsilon_{0}-\Sigma_{\sigma }^{0})+U\Sigma_{\overline{\sigma
}}^{1}},  \label{rgf}
\end{equation}
where $\widetilde{\epsilon }_{\sigma }=\epsilon_{0}+U+\Sigma
_{\sigma }^{0}+\Sigma_{\overline{\sigma }}^{3}$, $\Sigma_{\sigma
}^{L(R)0}=\int \frac{d\varepsilon'}{2\pi }\frac{\Gamma _{\sigma
}^{L(R)}(\varepsilon')}{\varepsilon-\varepsilon'+i\eta }$,
$\Sigma_{\overline{\sigma }}^{2}=\Sigma _{\overline{\sigma
}}^{3}-\Sigma_{\overline{\sigma }}^{1}$, $\Sigma _{\overline{\sigma
}}^{L(R)\lambda }=\int\frac{d\varepsilon'}{2\pi
}\Gamma_{\overline{\sigma }}^{L(R)}(\varepsilon')B\digamma
(\varepsilon')$, ($B=1$, $\lambda=3$; $B=f_{L(R)}$, $\lambda=1$),
where $\digamma (\varepsilon')=\frac{1}{\varepsilon -(2\epsilon
_{0}+U)+\varepsilon'+i\eta }+\frac{1}{\varepsilon
-\varepsilon'+i\eta }$, $\Gamma _{\sigma }^{L}(\varepsilon')=2\pi
\mathbf{\rho }_{L}(\varepsilon'+\sigma M)\left\vert t_{L}^{\sigma
}(\varepsilon')\right\vert ^{2}$, $\Gamma _{\uparrow }^{R}=\Gamma
_{\downarrow }^{R}=\Gamma^{R}(\varepsilon')=2\pi\rho
_{R}(\varepsilon')\left\vert t_{R}(\varepsilon')\right\vert^{2}$,
$\mathbf{\rho }_{L(R)}$ is density of state (DOS) of the left
(right) lead and $\Sigma_{\sigma }^{\gamma }=\Sigma_{\sigma
}^{L\gamma }+\Sigma_{\sigma }^{R\gamma }$ ($\gamma =1,2,3$). The
retard selfenergy can be derived from Dyson equation $\mathbf{\Sigma
}^{r}=(\mathbf{g}^{r})^{-1}-(\mathbf{G}^{r})^{-1}$, where
$\mathbf{g}^{r}$ is the retard GF of QD without coupling to the
leads but with Coulomb interaction. To get $\left\langle
n_{\overline{\sigma }}\right\rangle $, the selfconsistent
calculation must be preformed \cite{selfconsist}. And this procedure
needs lesser Green function which is subject to the Keldysh formula
$G^{<}=G^{r}\Sigma^{<}G^{a}$. The lesser self-energy is taken the
form as $\Sigma^{<}=\frac{1}{2}[\Sigma _{0}^{<}(\Sigma
_{0}^{r}-\Sigma_{0}^{a})^{-1}(\Sigma^{r}-\Sigma^{a})+(\Sigma
^{r}-\Sigma^{a})(\Sigma_{0}^{r}-\Sigma_{0}^{a})^{-1}\Sigma
_{0}^{<}]$ \cite{self}, where $\Sigma_{0}^{r(a,<)}$ are the
selfenergies of the noninteracting system while $\Sigma^{r(a,<)}$
are selfenergies with full interaction. In fact, one method without
solving $G^{<}$ and only with calculating the integral $\int
d\varepsilon G^{<}(\varepsilon)$ exactly has been developed to round
the calculation of the lesser Green function \cite{sun1}. However,
the approximation used here to derive the lesser Green function can
give a qualitatively correct results. We shall mention that
$\left\langle
d_{\overline{\sigma}}^{\dagger}d_{\sigma}\right\rangle$ in principle
tends to zero without spin flip scattering. We keep it here to avoid
any uncertainty which might be caused by self-consistent calculation
procedure and its value can be given by the self-consistent
calculation.

No losing generality, we shall do numerical calculations in the
limit $U\rightarrow\infty $. We use $\Gamma_{0}^{L}$ as the unit of
energy, which is defined in terms of the unpolarized parabolic
energy bands parameters, and $J_{0}=e\Gamma _{0}^{L}/\hbar $ as
current unit. We set $P=\frac{\rho_{\uparrow }^{L}-\rho_{\downarrow
}^{L}}{\rho_{\uparrow }^{L}+\rho_{\downarrow }^{L}}$,
$t_{qR(kL)}^{\uparrow }=t_{qR(kL)}^{\downarrow }=t_{R(L)}$, then
$\chi_{\uparrow }=\Gamma_{\uparrow
}^{L}/\Gamma_{0}^{L}=(\frac{2}{1+\chi^{2}})^{\frac{1}{2}}$, $\chi
_{\downarrow }=\Gamma_{\downarrow }^{L}/\Gamma_{0}^{L}=\chi\chi
_{\uparrow }$, where $\chi=\frac{1-P}{1+P}$. Let
$\chi_{R}=\Gamma^{R}/\Gamma_{0}^{L}=\alpha $. The left lead is FM
and the right lead is SC, and the $\Gamma^{L(R)}$ is in proportion
to the DOS of the left (right) lead. So we may estimate the $\alpha$
will be between $10^{-4}\rightarrow 10^{-3}$ (for the right SC lead,
we use 3D DOS rather than using 2D DOS to avoid the complexity). But
if the tunnelling matrix $t_{(L)R}$ can be tuned to be different,
the parameter $\alpha$ may be tuned till $1$ (this case is
considered in Fig. 4). To get the retard Green function, selfenergy
$\Sigma_{\sigma}^{1}$ will be calculated analytically as \cite{s1}
\begin{equation}
\Sigma _{\sigma }^{L1}(\Sigma^{R1})=\frac{\Gamma_{\sigma
}^{L}(\Gamma ^{R})}{2\pi }\{\frac{1}{2}\ln \frac{\varepsilon
_{1}\varepsilon _{2}}{(2\pi iT)^{2}}-\psi (z)-\frac{i}{2}\pi \},
\label{sel1}
\end{equation}
where $\varepsilon _{1}=\varepsilon-\mu _{L(R)}-D$, $\varepsilon
_{2}=\varepsilon-\mu_{L(R)}+D$, $z=\frac{1}{2}+\frac{\varepsilon-\mu
_{L(R)}}{2\pi iT}$, $D$ is the half bandwidth, and we set it as 1500
in this letter, $T$ is temperature.
%For symmetry case (i.e. $\alpha =1$) in Fig. (2a), the first character is the
%splitting of $\rho _{\uparrow }(\varepsilon )$ and $\rho
%_{\downarrow }(\varepsilon )$.

%%%%%%%%%%%%%%%%%%%%%%%%%%%%%%%%%%%%%%%%%%%%%%%%%%%%%%%%%%
\begin{figure}[tbph]
\centering \includegraphics[width =13.5 cm, height=11 cm]{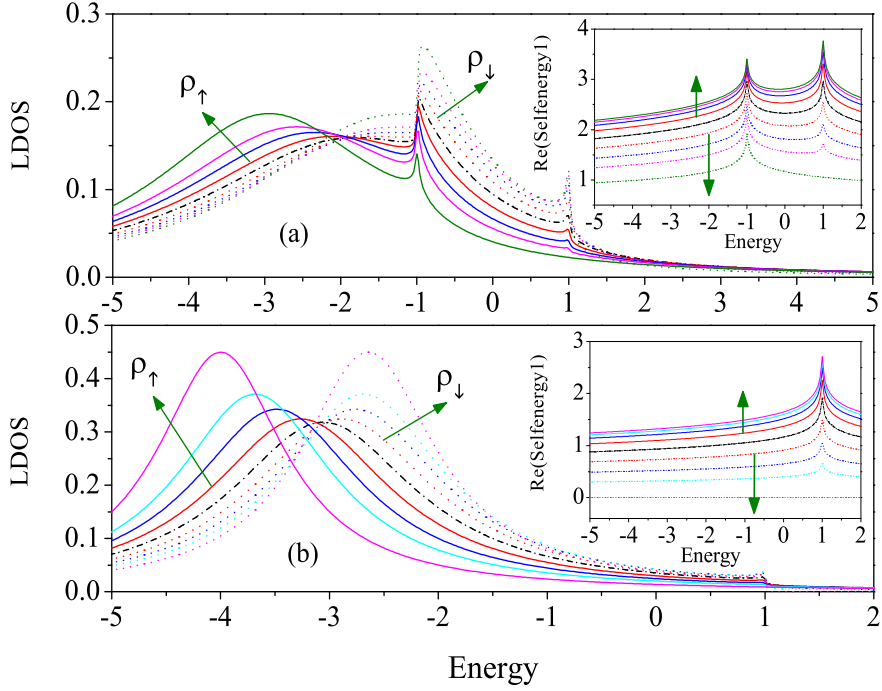}
\caption{(color online). In the four graphs, dash-dot lines are
generated at $P=0$, and solid (dot) lines for $\rho_{\uparrow }$
($\rho _{\downarrow }$) or $\Re\Sigma_{\uparrow }^{1}$ ($\Re\Sigma
_{\downarrow }^{1}$ ). Along the directions of the arrows, the lines
are sequently corresponding to $P=0.2,$ $0.4,$ $0.6,$ and $1$. LDOS
vs. energy in (a) and (b), $\Re\Sigma_{\sigma }^{1}$ in insets.
$\alpha =1$ in (a) and its inset; $\alpha=0.001$ in (b) and its
inset. The other parameters are $T=0.01$, $\varepsilon_{0}=-4$, and
$V=2$ ($\varepsilon_{0}$ and $V$ are the same in all figures). }
\label{fig2}
\end{figure}
%%%%%%%%%%%%%%%%%%%%%%%%%%%%%%%%%%%%%%%%%%%%%%%%%%%%%%%%
Local density of state (LDOS) \cite{ldos} $\rho_{\sigma}(\varepsilon
)$ and $\Re\Sigma _{\sigma }^{1}$ in QD vs. energy for different
polarization $P$ are shown in Fig. 2. It can be found that
$\rho_{\uparrow }(\varepsilon )$ and $\rho _{\downarrow
}(\varepsilon )$ are split because of FM lead. The main peaks of
LDOS exist at the resonant energy $\varepsilon =\varepsilon
_{0\sigma }$, where $\varepsilon _{0\sigma }=\varepsilon
_{0}+\Re\Sigma _{\overline{\sigma }}^{1}$ indicates that the
original spin-independent energy level is modified as spin-dependent
energy levels because of Coulomb interaction on QD and the
polarization $P$. $\Re\Sigma _{\overline{\sigma }}^{1}$ is
proportional to the tunnelling rate $\Gamma _{\overline{\sigma
}}^{L}$ which depends on $P$. $\Re\Sigma _{\uparrow }^{1}$
($\Re\Sigma _{\downarrow }^{1}$) increases (decreases) with $P$ and
show two resonances at $-V/2$ and $V/2$ except for $\Re\Sigma
_{\downarrow }^{1}$ at $P=1$ as shown in the inset of Fig. (2a).
This increasing (decreasing) gives rise to the shift of the main
peak of $\rho_{\uparrow }(\varepsilon )$ ($\rho _{\downarrow
}(\varepsilon )$) towards to the lower (higher) energy.

Kondo resonances (KRs) of $\rho_{\uparrow }(\varepsilon )$ and $\rho
_{\downarrow }(\varepsilon )$ existing about at $\mu_{L}=V/2$ and
$\mu _{R}=-V/2$ shown in Fig. (2a) are consequences of
nonequilibrium effect \cite{meir}. i) With increasing $P$, the
magnitude of KRs for down (up) spin component is higher (lower).
While the peak of $\Re\Sigma _{\downarrow }^{1}$ disappears at $V/2$
and $P=1$ in the inset of Fig. (2a), the corresponding KR of $\rho
_{\uparrow }(\varepsilon )$ disappears also. But the other KRs of
$\rho _{\uparrow }(\varepsilon )$ bounded at $-V/2$ remain. The
reason is there are no itinerant spin-down electrons in the FM layer
for formation of spin singlet with the electrons on QD now. ii) The
positions of spin-up KRs move to lower energy and the spin-down ones
move to the opposite direction. Spin-dependent Kondo effect was
firstly investigated in FM-QD-FM system in Ref. \cite{sergueev}.
Then further theoretical \cite{martinek} and experimental \cite{pas}
investigations are evaluated to show the splitting of the Kondo
resonances. However it has no influences on the spin filter effect
which is mainly investigated in a temperature scale much above the
Kondo temperature.

% which is consistent with recent
%theoretical prediction for FM-QD-FM system \cite{martinek} and the
%recent experimental result \cite{pas}.

Higher order cotunneling processes \cite{sasakl} depicted in Fig.
(1b) account for the formation of these KRs. Initially, an up-spin
electron occupies the QD, it can jump to the left (right) lead at a
time scale $\hbar /(\mu _{L(R)}-\varepsilon _{0})$. Almost at the
same time, a down-spin electron of the right (left) lead can jump
into the QD. Then the final state is a spin-flip state. A large
number of coherent superpositions of these events will give rise to
KRs at Fermi levels. The conduction electrons tend to screen the
nonzero spin on QD such that a many-body spin singlet state forms.
This process can transfer charges from one lead to the other and
then the KR may enhance the conductance or current \cite{ng}.
Another contributing process is that a spin occupying the QD jumps
into one lead and almost at the same time an opposite spin in the
same lead tunnels into QD, which can be clearly seen from the Eq.
(239) in Ref. \cite{platero} in which the Hamiltonian consisting the
QD and leads is transformed into a Hamiltonian similar to the
conventional Kondo Hamiltonian via Schrieffer-Wolff transformation
\cite{sw}. There $J_{LL}$ and $J_{RR}$ describe the coupling of the
local spin on QD and the spins of itinerant electrons in the left or
right leads respectively. This process doesn't transfer charge from
one lead to the other.

When $\alpha $ is very small (for example $\alpha=0.001$ in Fig.
(2b) and its inset), $\Sigma^{R1}$ contributes little to
$\Re\Sigma_{\sigma }^{1}$. The channel of formation KRs between QD
and the right lead is suppressed. So in Fig. (2b) and its inset, the
peaks about at $-V/2$ all disappear, but the KRs about at $V/2$ are
still present. The spin-splitting of $\rho _{\uparrow
}(\varepsilon)$ and $\rho_{\downarrow }(\varepsilon )$ remains,
giving rise to the spin filter effect described in the following.

To investigate the spin-filter effect, we shall calculate the
current through this structure. By using the nonequilibrium Green
function technique \cite{negf}, the steady current with up (down)
spin in unit $J_{0}$ is
\begin{equation}
J_{\uparrow (\downarrow )}/J_{0}=\Delta _{1(2)}\int d\varepsilon \lbrack
f_{L}(\varepsilon )-f_{R}(\varepsilon )]\rho _{d\uparrow (d\downarrow
)}(\varepsilon ),  \label{jud}
\end{equation}%
where $\Delta _{1}=\frac{\chi _{\uparrow }\alpha }{\chi _{\uparrow
}+\alpha } $ and $\Delta _{2}=\frac{\chi _{\downarrow }\alpha }{\chi
_{\downarrow }+\alpha }$. When a gate voltage is applied, we set new
energy level on QD is $\varepsilon _{0}'=\varepsilon _{0}+V_{g}$.
The spin polarization of current is defined as
$P_{out}=\frac{J_{\uparrow }-J_{\downarrow }}{J_{\uparrow
}+J_{\downarrow }}$, which is not the polarization of the DOS of the
left lead. When $P=0$, i.e. the injector is spin independent,
$\Delta _{1}=\Delta _{2}$ and $P_{out}=0$. When $P=1$, i.e. the
injector is fully polarized (for example half-metal material),
$\Delta _{2}=0$ and $P_{out}=1$. When $0<P<1,$ at the limit $\alpha
\rightarrow 0$, $\Delta _{1}\approx \Delta _{2}\approx \alpha $,
then $P_{out}$ just depends on the difference of the LDOS of QD for
different spins.

%%%%%%%%%%%%%%%%%%%%%%%%%%%%%%%%%%%%%%%%%%%%%%%%%%%%%%%%%%
\begin{figure}[tbph]
\centering \includegraphics[width =10 cm, height=11 cm]{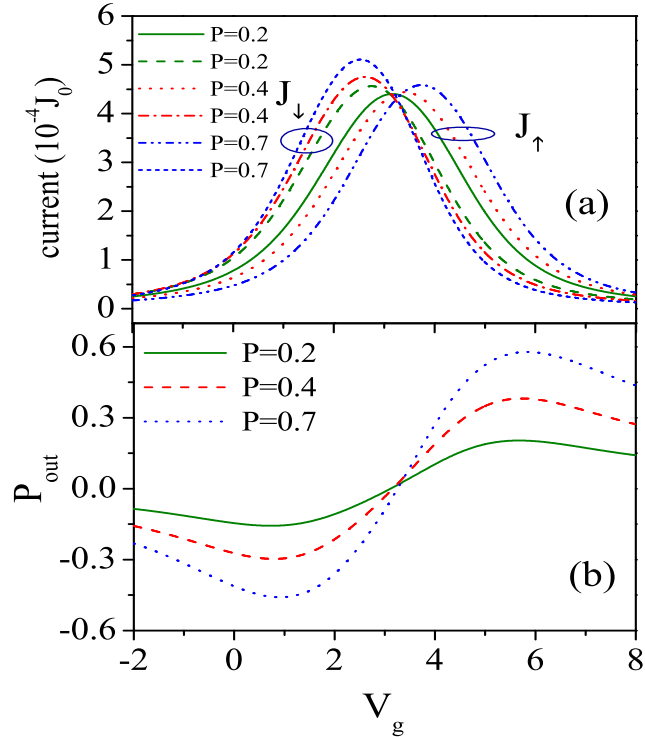}
\caption{(color online). The $V_{g}$ dependence of current in (a)
and $P_{out}$ in (b). $\alpha =0.001$ and $T=0.5$.} \label{fig3}
\end{figure}
%%%%%%%%%%%%%%%%%%%%%%%%%%%%%%%%%%%%%%%%%%%%%%%%%%%%%%%%

Quite usually, devices are operated in room temperature which is
much higher than Kondo temperature, and KRs disappear. The $V_{g}$
dependence of $J_{\uparrow }$ and $J_{\downarrow }$ are presented in
Fig. (3a). It is noted that currents have resonant peaks which are
split under nonzero $P$ because of $\varepsilon _{0\uparrow }\neq
\varepsilon _{0\downarrow }$. And the splitting becomes larger with
increasing $P$. It can be understood that when $\varepsilon
_{0\uparrow (\downarrow )}$ is in the energy range $[-V/2,V/2]$ as
shown in Fig. (1c) and (1d), resonant tunnelling occurs and a
resonant peak of $J_{\uparrow (\downarrow )}$ present. When
$\varepsilon _{0\uparrow (\downarrow )}$ is out of this range,
current is suppressed. We call this range as resonant window (RW).
As $\varepsilon _{0\downarrow }>\varepsilon _{0\uparrow }$,
$\varepsilon _{0\downarrow }$ first enters into the RW with
increasing $V_{g}$ as shown in Fig. (1c), now $J_{\downarrow }$ is
on-resonant and $J_{\uparrow }$ is off-resonant. Increasing $V_{g}$
further, $\varepsilon_{0\uparrow }$ enters into the RW as shown in
Fig. (1d), and the case is opposite to the former. When
$V_{g}>V_{g0}$ (we set $J_{\uparrow }=J_{\downarrow }$ at $V_{g0}$),
$P_{out}>0$; $P_{out}$ first increases and then decreases with
$V_{g}$. When $V_{g}<V_{g0}$, $P_{out}<0$; $P_{out}$ also first
increases and then decreases with $V_{g}$ as shown in Fig. (3b).
Even $\alpha $ is very small, there is still a large $P_{out}$. For
example, when $P=0.7$, the peak magnitude of $P_{out}$ is 0.58
($V_{g}=5.83$), and the peak magnitude is enhanced by increasing
$P$. $P_{out}$ varies from negative to positive with increasing gate
voltage, which means the spin filter effect can be controlled by
tuning the gate voltage.
%%%%%%%%%%%%%%%%%%%%%%%%%%%%%%%%%%%%%%%%%%%%%%%%%%%%%%%%%%
\begin{figure}[tbph]
\centering \includegraphics[width =10 cm, height=10 cm]{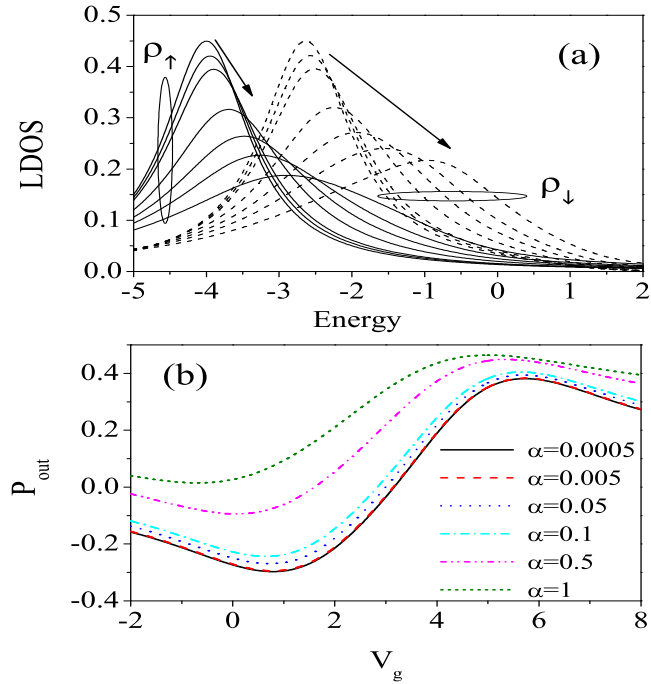}
\caption{(color online). LDOS vs. energy in (a) and $V_{g}$
dependence of $P_{out}$ in (b) for different $\alpha $. In (a) along
the directions of arrows the lines correspond to $\alpha =0.001,$
$0.05$, $0.1$, $0.3$, $0.5$, $0.7$, and $1$. The other parameters
are $P=0.4,$ $T=0.5$.} \label{fig4}
\end{figure}
%%%%%%%%%%%%%%%%%%%%%%%%%%%%%%%%%%%%%%%%%%%%%%%%%%%%%%%%

LDOS vs. energy in Fig. (4a) and $V_{g}$ dependence of $P_{out}$ in
Fig. (4b) for different $\alpha $. $\rho _{\uparrow }$ and $\rho
_{\downarrow }$ become lower and fatter with increasing $\alpha $.
It means the local electrons on QD tend to be nonlocal and tunnel to
the right lead. And the main peaks are shifted towards the right
direction shown in Fig. (4a). It is found that negative $P_{out}$ is
reduced and even becomes positive with increasing $\alpha $ in Fig.
(4b). On the other hand, the positive $P_{out}$ will be enhanced
with $\alpha $. For example, when $\alpha =1$, there is no negative
$P_{out}$. But the maximum of $P_{out}$ is enhanced to give 0.464
for $\alpha =1$ and $P=0.4$.

The effect predicted here is the consequence of well-defined
spin-dependent energy levels of QD. So spin relaxation in QD may
reduce the effect and we may estimate its order. For comparison, in
Ref. \cite{gruber}, spin relaxation time (SRT) is shorter in ZnMnSe
layer because of the spin-dependent scatterings in it. However, SRT
is much longer in our case. Firstly, it is because the QD is formed
in the nonmagnetic semiconductor quantum well, spin-dependent
scatterings are sparse. Secondly, the zero dimensionality of
electron states in QDs leads to a significant suppression of the
most effective 2D spin-flip mechanisms \cite{kha}, and the electron
spin states in QDs are expected to be very stable. Recent electrical
transport measurements of relaxation between spin triplet and
singlet states confined in a VQD give relaxation time $>200 \mu s$
at $T\leq 0.5$ K \cite{fuj}. Finally, we estimate the transit time.
For a typical value $\Gamma =150$ $\mu V$ \cite{sasakl} ($\Gamma $
can be changed by changing the barrier thickness \cite{gueret}), the
estimated transit time is about $5$ $ps$. So it seems reasonable to
assume that the spin relaxation on QD has little effect in this
model.

In Ref. \cite{gruber}, the spin dependent energy levels are induced
by Zeeman splitting under an external magnetic field in magnetic
semiconductor ZnMnSe quantum well. While in this letter the
tunnelling rates for up and down spins are split because of the
splitting of DOS of FM. This splitting likes an effective magnetic
field (EMF) but much stronger than conventional magnetic field, even
reach $50\sim 70$ T \cite{ralph}, leading to the well defined
spin-dependent energy levels on QD. Further an upper limit on the
local magnetic field (LMF) which is generated by FM lead in QD is
estimated to be 0.6 T for Ni \cite{hanson}. It seems reasonable to
neglect this LMF.

In summary, selective and large polarization of current injected
into semiconductor is predicted in Ferromagnet /Quantum Dot
/semiconductor system by varying the gate voltage above the Kondo
temperature. A FM layer is used as a spin source and electrons
tunnel through a QD into SC. Spin-dependent Kondo effect is revealed
below Kondo temperature. KRs for up spin state is suppressed with
$P$. While the down one is enhanced. The KR for up spin is disappear
at $P=1$. With increasing the gate voltage, the polarization of
current varies from negative to positive, which means spin filter
effect can be controlled by gate voltage. A large efficient spin
injection can be obtained.

This work was supported by the Natural Science Foundation of China
(Grant Nos. 10574076, 10447118), and by the Program of Basic
Research Development of China (Grant No. 2006CB921500).


\begin{thebibliography}{}
\bibitem{assum} It means that correlation function which invovle unlike
spin-indices, invovle different operators of the left and the right
leads or both creat operators (annihlation operators) equal to zero,
and factorize the correlation functions with like spin
$\langle\langle a_{k'\overline{\sigma }}^{\dagger
}a_{k''\overline{\sigma }}d_{\sigma }|d_{\sigma'}^{\dagger }\rangle
\rangle ^{r}\approx f(\varepsilon _{k'\overline{\sigma }}^{L})\delta
_{k'k''}\langle \langle d_{\sigma }|d_{\sigma'}^{\dagger }\rangle
\rangle^{r}$.
\bibitem{selfconsist} $\left\langle n_{\sigma }\right\rangle =\Im\int
\frac{d\varepsilon }{2\pi }\langle \langle d_{\sigma }|d_{\sigma
}^{\dagger }\rangle \rangle^{<}$, $\left\langle d_{\sigma }^{\dagger
}d_{\overline{\sigma }}\right\rangle =\Im\int \frac{d\varepsilon
}{2\pi }\langle \langle d_{\overline{\sigma }}|d_{\sigma }^{\dagger
}\rangle\rangle^{<}$, where $\langle \langle a|b\rangle \rangle^{<}$
means lesser GF derived from the returded GF by using Keldysh
formula. Given these two initial values, then substitute them into
the retard Green function and Keldysh formula to get them for the
second time. This procedure continues to give a constant
$\left\langle n_{\sigma }\right\rangle $ and $\left\langle d_{\sigma
}^{\dagger }d_{\overline{\sigma }}\right\rangle $. The retard Green
function and lesser Green function can be obtained.
\bibitem{ldos} The local density of states (LDOS) is defined as $\rho
_{\sigma }(\varepsilon )=-\frac{1}{\pi }\Im G_{d}^{\sigma \sigma
,r}(\varepsilon )$.
\end{thebibliography}
\end{document}